\documentclass[nofootinbib,prd]{revtex4}%
\usepackage{amsmath}
\usepackage{amsfonts}
\usepackage{amssymb}
\usepackage{graphicx}%
\begin{document}
\title{Wormholes or Gravastars?}
\author{Remo Garattini}
\email{Remo.Garattini@unibg.it}
\affiliation{Universit\`{a} degli Studi di Bergamo, Facolt\`{a} di Ingegneria,}
\affiliation{Viale Marconi 5, 24044 Dalmine (Bergamo) Italy}
\affiliation{and I.N.F.N. - sezione di Milano, Milan, Italy.}

\begin{abstract}
The one loop effective action in a Schwarzschild background is here used to
compute the Zero Point Energy (ZPE) which is compared to the same one
generated by a gravastar. We find that only when we set up a difference
between ZPE in these different background we can have an indication on which
configuration is favored. Such a ZPE difference represents the Casimir energy.
It is shown that the expression of the ZPE is equivalent to the one computed
by means of a variational approach. To handle with ZPE divergences, we use the
zeta function regularization. A renormalization procedure to remove the
infinities together with a renormalization group equation is introduced. We
find that the final configuration is dependent on the ratio between the radius
of the wormhole augmented by the "brick wall" and the radius of the gravastar.
\end{abstract}
\maketitle

\section{Introduction}

Black holes are well accepted astrophysical objects by scientific community.
The simplest example of a black hole is the spherically symmetric vacuum
solution of the Einstein field equations,%
\begin{equation}
ds^{2}=-\left(  1-\frac{2MG}{r}\right)  dt^{2}+\frac{dr^{2}}{1-\frac{2MG}{r}%
}+r^{2}\left(  d\theta^{2}+\sin^{2}\theta d\phi^{2}\right)  , \label{Schw}%
\end{equation}
known as the Schwarzschild solution. Despite of many theoretical and
observational successes, a number of paradoxical problems connected to black
holes also exist\cite{Wald}, which frequently motivate authors to look for
other alternatives, in which the endpoints of gravitational collapse are
massive stars without horizons. In 2001, Mazur and Mottola\cite{MM} proposed
an alternative model to black hole as a different final state of a
gravitational collapse. In this model, the strong gravitational forces induce
a vacuum rearrangement in such a way to avoid a classical event horizon. A
phase transition is associated to the quantum gravitational vacuum together
with a topology change. Such a model has been termed \textit{gravastar}
(\textit{grav}itational \textit{vac}uum \textit{star}) and it consists of
three different regions with three different Eqs. of state%
\begin{equation}%
\begin{array}
[c]{rrr}%
I.\qquad\mathrm{Interior:} & 0\leq r<r_{1} & \rho=-p,\\
II.\qquad\mathrm{Thin}\text{ }\mathrm{Shell:} & r_{1}<r<r_{2} & \rho=+p,\\
III.\qquad\mathrm{Exterior:} & r_{2}<r & \rho=p=0.
\end{array}
\label{GStar}%
\end{equation}
At the interfaces $r=r_{1}$ and $r=r_{2}$, we require the metric coefficients
to be continuous, although the associated first derivatives must be
discontinuous. Globally the metric can be cast into a form very close to the
Schwarzschild line element $\left(  \ref{Schw}\right)  $%
\begin{equation}
ds^{2}=-N^{2}\left(  r\right)  dt^{2}+\frac{dr^{2}}{1-\frac{b\left(  r\right)
}{r}}+r^{2}\left(  d\theta^{2}+\sin^{2}\theta d\phi^{2}\right)  ,
\label{metric}%
\end{equation}
where%
\begin{equation}
N^{2}\left(  r\right)  =1-\frac{\Lambda_{dS}}{3}r^{2}\qquad\mathrm{and\qquad
}b\left(  r\right)  =\frac{\Lambda_{dS}}{3}r^{3}%
\end{equation}
for the interior region of a static de Sitter metric, while%
\begin{equation}
N^{2}\left(  r\right)  =1-\frac{2MG}{r}\qquad\mathrm{and\qquad}b\left(
r\right)  =2MG
\end{equation}
for the exterior region described by a Schwarzschild spacetime. The
intermediate region is represented by a thin shell endowed with a Minkowski
metric. However, the thin shell is by no means necessary to obtain a
gravastar. Indeed, DeBenedictis et al. \cite{DBHIKV} proved that the shell
region can be eliminated and the de Sitter spacetime can be directly joined to
the Schwarzschild metric. This picture was also considered by
Dymnikova\cite{Dymnikova} without invoking the term \textquotedblleft%
\textit{gravastar}\textquotedblright. From the asymptotic point of view, a
gravastar and a black hole share the same Arnowitt-Deser-Misner
($\mathcal{ADM}$) mass\cite{ADM}, therefore at very large distances they
appear the same object to an observer. One may wonder how can we distinguish a
black hole from a gravastar. Different proposal have been considered. Harko et
al.\cite{HKL} suggested to compare the thermodynamical and electromagnetic
properties of the accretion disk around slowly rotating black holes and
gravastars. Another proposal comes from Chirenti and Rezzolla\cite{CR}, where
the use of axial perturbations and the analysis of Quasi Normal Modes seems to
show how to distinguish a black hole from a gravastar. Cardoso et
al.\cite{CPCC} discussed the ergoregion instability produced by rapidly
spinning compact objects such as gravastars and boson stars with the results
that ultra-compact objects with large rotation are black holes. On the other
side, Chirenti and Rezzolla\cite{CR1} found that stable models can be
constructed also with $J/M^{2}\sim1$, where $J$ and $M$ are the angular
momentum and mass of the gravastar, respectively. Note that the comparison
between a gravastar and a black hole is at the classical level, without any
quantum contribution which should also be the source of the desired phase
transition to make a gravastar. Note also that, in principle there exists
another source of ambiguity: indeed besides a gravastar and a black hole, the
exterior metric can be associated to a wormhole. In this paper, we would like
to compare a gravastar to a wormhole. From the asymptotic point of view even
the wormhole shares the same $\mathcal{ADM}$ mass. Hence, it should be very
important the comparison of such objects invoking the Zero Point Energy (ZPE)
contribution. However, it is well known that every form of ZPE also
contributes to an induced cosmological constant. If we indicate with
$E_{0}^{GS}$ the gravastar ZPE and with $E_{0}^{W}$, the wormhole ZPE, in
principle one can discuss the following inequalities%
\begin{equation}
\left(  E_{0}^{W}-E_{0}^{GS}\right)  \gtreqless0 \label{ineq}%
\end{equation}
establishing which geometry is energetically favored compared to the other
one. Essentially, this is a Casimir-like calculation and lower the ZPE, the
more stable is the final configuration. This inequality is also related to the
decay probability per unit volume and time $\Gamma$, which is defined as%
\begin{equation}
\Gamma=A\exp\left(  -I_{cl}\right)  =\exp\left(  -I_{g}\left[  \bar{g}_{\mu
\nu}\right]  \right)  \int\mathcal{D}h_{\mu\nu}\exp\left(  -I_{g}^{\left(
2\right)  }\left[  h_{\mu\nu}\right]  \right)  ,
\end{equation}
for an Euclidean time. Indeed, at least a tree level, we find that%
\begin{equation}
\Gamma\simeq\exp\left(  -I_{cl}\right)  =\exp\left(  -\left(  E_{0}^{GS}%
-E_{0}^{W}\right)  \Delta\tau\right)  ,
\end{equation}
where $\Delta\tau$ is the Euclidean time interval. Basically, the first step
is the evaluation of the following expectation value%
\begin{equation}
\left\langle T_{\mu\nu}\right\rangle =\frac{2}{\sqrt{-g}}\frac{\delta
\Gamma_{g}}{\delta g^{\mu\nu}}, \label{eff}%
\end{equation}
provided one can compute the effective action related to $S_{g}$. To do
calculations in practice, we fix our attention to the standard Einstein action
without matter fields and with a cosmological term%
\begin{equation}
S=S_{g}+S_{C.C.}%
\end{equation}
where%
\begin{equation}
S_{g}=\frac{1}{16\pi G}\int d^{4}x\sqrt{-g}R,\qquad S_{C.C.}=-\frac{\Lambda
}{8\pi G}\int d^{4}x\sqrt{-g}.
\end{equation}
The least action principle leads to the Einstein's fields equations with a
cosmological term. Nevertheless, never forbids to consider the cosmological
term as the desired induced quantity by ZPE. Therefore, Eq.$\left(
\ref{eff}\right)  $ must be modified into%
\begin{equation}
\left\langle T_{\mu\nu}^{C.C.}\right\rangle =-\frac{2}{\sqrt{-g}}\frac
{\delta\Gamma_{g}}{\delta g^{\mu\nu}}. \label{effCC}%
\end{equation}
If we define the path integral%
\begin{equation}
Z=\int\mathcal{D}\left[  g_{\mu\nu}\right]  \exp iS_{g}\left[  g_{\mu\nu
}\right]  \label{p01}%
\end{equation}
and we consider a gravitational field of the form%
\begin{equation}
g_{\mu\nu}=\bar{g}_{\mu\nu}+h_{\mu\nu}, \label{pert4}%
\end{equation}
then Eq.$\left(  \ref{p01}\right)  $ becomes
\begin{equation}
\int\mathcal{D}g_{\mu\nu}\exp iS_{g}\left[  g_{\mu\nu}\right]  =\exp
iS_{g}\left[  \bar{g}_{\mu\nu}\right]  Z_{2}=\exp iS_{g}\left[  \bar{g}%
_{\mu\nu}\right]  \int\mathcal{D}h_{\mu\nu}\exp iS_{g}^{\left(  2\right)
}\left[  h_{\mu\nu}\right]  , \label{p02}%
\end{equation}
where we have assumed that the background $\bar{g}_{\mu\nu}$ is a solution of
the Einstein field equations and%
\begin{equation}
S_{g}^{\left(  2\right)  }\left[  h_{\mu\nu}\right]  =\frac{1}{2\kappa}%
\int_{\mathcal{M}}d^{4}x\sqrt{-g}h_{\mu\rho}O^{\mu\rho\sigma\nu}\!{}%
h_{\sigma\nu} \label{quadr}%
\end{equation}
with $\kappa=8\pi G$. $O^{\mu\rho\sigma\nu}$ is a symmetric tensor operator
with%
\begin{equation}
O^{\mu\rho\sigma\nu}=\frac{\delta^{2}S\left[  \bar{g}_{\mu\nu}\right]
}{\delta h^{\mu\rho}\delta h^{\sigma\nu}} \label{Ope}%
\end{equation}
and $h_{\sigma\nu}$ is the quantum fluctuation with respect to the background
$\bar{g}_{\sigma\nu}$. After some integration by parts, Eq.$\left(
\ref{quadr}\right)  $ becomes%
\[
S^{\left(  2\right)  }=\frac{1}{2\kappa}\int_{\mathcal{M}}d^{4}x\sqrt
{-g}\left[  -\frac{1}{4}h^{\mu\nu}\left(  \bigtriangleup_{L\!}\!{}h\right)
_{\mu\nu}+\frac{1}{4}h\bigtriangleup\!{}h+h^{\mu\rho}R_{\rho\nu}h_{\mu}^{\nu
}-\frac{1}{2}h^{\mu\nu}\!{}h_{\mu;\alpha;\nu}^{\alpha}\right.
\]%
\begin{equation}
\left.  -\frac{1}{2}hR_{\alpha\beta}h^{\alpha\beta}+\frac{1}{2}h\!{}%
h_{;\mu;\nu}^{\mu\nu}+R\left(  -\frac{1}{4}h^{\mu\nu}h_{\mu\nu}+\frac{1}%
{8}h^{2}\right)  \right]  . \label{S24d}%
\end{equation}
$\bigtriangleup_{L}$stands for the Lichnerowicz operator defined by%
\begin{equation}
\left(  \bigtriangleup_{L}h\right)  _{\mu\nu}=-\nabla^{a}\nabla_{a}h_{\mu\nu
}-2R_{\mu\alpha\nu\beta}h^{\alpha\beta}+R_{\mu\alpha}h_{\nu}^{\alpha}%
+R_{\nu\alpha}h_{\mu}^{\alpha}=\bigtriangleup h_{\mu\nu}-2R_{\mu\alpha\nu
\beta}h^{\alpha\beta}+R_{\mu\alpha}h_{\nu}^{\alpha}+R_{\nu\alpha}h_{\mu
}^{\alpha}, \label{dl4d}%
\end{equation}
where we have introduced the positive definite differential operator%
\begin{equation}
\bigtriangleup=-\nabla^{a}\nabla_{a}. \label{Lapl}%
\end{equation}
$S^{\left(  2\right)  }$ simplifies considerably when we are on shell, namely
$R_{\alpha\beta}=0$. Nevertheless, for future purposes, it is convenient
keeping such terms in the expression of the Lichnerowicz operator and in
Eq.$\left(  \ref{S24d}\right)  $. To extract physical informations from
expression $\left(  \ref{p02}\right)  $, we need an orthogonal decomposition
which is equivalent to the Faddeev-Popov procedure, at least to one loop. From
Appendix \ref{AppA}, we obtain\cite{GP,CD}%
\begin{equation}
\Gamma_{1-loop}=\frac{i}{2}\left[  Tr\ln\bigtriangleup_{L\!}^{\bot}%
\!-Tr\ln\triangle_{V^{\bot}}\right]  =\frac{i}{2}\int_{\mathcal{M}}d^{4}%
x\sqrt{-g}\left[  \int\frac{d^{4}k}{\left(  2\pi\right)  ^{4}}\ln\lambda
_{TT}^{2}-\int\frac{d^{4}k}{\left(  2\pi\right)  ^{4}}\ln\lambda_{V^{\bot}%
}^{2}\right]  , \label{G1L}%
\end{equation}
where $\lambda_{TT}^{2}$ and $\lambda_{V^{\bot}}^{2}$ are the eigenvalues of
the Lichnerowicz operator for TT tensors and the transverse vector operator
respectively. With the help of Eq.$\left(  \ref{G1L}\right)  $, Eq.$\left(
\ref{effCC}\right)  $ simply becomes%
\begin{equation}
\left\langle T_{\mu\nu}^{C.C.}\right\rangle =-\frac{i}{2}g_{\mu\nu}\left[
\int\frac{d^{4}k}{\left(  2\pi\right)  ^{4}}\ln\lambda_{TT}^{2}-\int
\frac{d^{4}k}{\left(  2\pi\right)  ^{4}}\ln\lambda_{V^{\bot}}^{2}\right]
\label{Tmn}%
\end{equation}
and if we identify%
\begin{equation}
\frac{\Lambda}{8\pi G}=-\frac{i}{2}\left[  \int\frac{d^{4}k}{\left(
2\pi\right)  ^{4}}\ln\lambda_{TT}^{2}-\int\frac{d^{4}k}{\left(  2\pi\right)
^{4}}\ln\lambda_{V^{\bot}}^{2}\right]  , \label{LoverG}%
\end{equation}
we can interpret the cosmological constant as induced by quantum fluctuations
of the gravitation field itself. Therefore, it is clear that inequality
$\left(  \ref{ineq}\right)  $ can be directly measured by the induced
cosmological quantity of Eq.$\left(  \ref{LoverG}\right)  $. Moreover this
identification will be useful for the removal of divergences. A first
observation about Eq.$\left(  \ref{G1L}\right)  $ is in order. Note that
nothing has been said regarding the famous conformal factor problem. From this
point of view, we adopt the approach of Mazur and Mottola\cite{MM1} in
decomposing the gravitational perturbation. The super-metric free parameter
\textquotedblleft$C$\textquotedblright\ of Eq.$\left(  \ref{superm}\right)  $
leaves us the freedom to select the correct range in such a way the functional
integration be convergent. Coming back to the Lichnerowicz operator $\left(
\ref{dl4d}\right)  $, one immediately recognize that finding the eigenvalues
is not a trivial task in general. It is therefore necessary to adopt a
convenient choice to manage Eq.$\left(  \ref{Tmn}\right)  $. In a previous
work we approached the cosmological constant problem with the help of the
Wheeler-DeWitt Equation cast in the form of a Sturm-Liouville
problem\cite{Remo}. Essentially the cosmological constant is reinterpreted as
an eigenvalue, calculated in a Hamiltonian formalism breaking the covariance
of space-time. In the next section, we adopt the same strategy provided one
looks at the true degrees of freedom. The paper is organized as follows: in
section \ref{p1}, we reduce the effective action by restricting the modes of
the perturbation, in section \ref{p2}, we evaluate the functional determinants
by means of a W.K.B. method, in section \ref{p3} we compute the ZPE energy for
the gravastar and the wormhole respectively and we compare them. Finally, in
section \ref{p4} we conclude.

\section{Reducing the one loop effective action in 3+1 dimensions}

\label{p1}How the Lichnerowicz operator decomposes in 3+1 dimensions it
depends on the way one separates space from time. The $\mathcal{ADM}$
variables offer a valid example of such a decomposition. In terms of these
variables, the metric background written in Eq.$\left(  \ref{metric}\right)  $
becomes%
\begin{equation}
ds^{2}=-N^{2}dt^{2}+g_{ij}\left(  N^{i}dt+dx^{i}\right)  \left(
N^{j}dt+dx^{j}\right)  .\label{ds2}%
\end{equation}
We recognize that the \textit{lapse function} $N$ is invariant and the
\textit{shift function} $N_{i}$ is absent. To have an effective reduction of
the modes, we consider perturbations of the gravitational field on the
hypersurface $\Sigma\subset\mathcal{M}$. This means that we are
\textquotedblleft\textit{freezing\textquotedblright\ }the perturbation of the
lapse and the shift functions respectively. In summary,%
\begin{equation}
\left\{
\begin{array}
[c]{c}%
g_{ij}\longrightarrow\bar{g}_{ij}+h_{ij}\\
N\longrightarrow N\\
N_{i}\longrightarrow0
\end{array}
\right.  \label{mpert}%
\end{equation}
corresponding to a restriction of the modes we are looking at. Choice $\left(
\ref{mpert}\right)  $ is equivalent to set
\begin{equation}
\!{}h_{0\mu}=0\qquad\left(  \mu=0,\ldots,3\right)  \label{Red}%
\end{equation}
in Eq.$\left(  \ref{S24d}\right)  $, which can be reduced to%
\begin{equation}
S^{\left(  2\right)  }=\frac{1}{2\kappa}\int_{\mathcal{M}}d^{4}x\sqrt
{-g}\left[  -\frac{1}{4}h^{\bot\ ij}\left(  \bigtriangleup_{L\!}\!{}h\right)
_{ij}^{\bot}+\frac{3}{32}\sigma\bigtriangleup\!{}\sigma\right]
.\label{SepS2a}%
\end{equation}
Note that the modes we have eliminated satisfy the transverse traceless
condition. The remaining modes are described only by spatial indices which are
raised and lowered using $\bar{g}_{ij}$ and $\bar{g}^{ij}$. Christoffel
symbols and Riemann tensor are entirely constructed with the help of the three
dimensional background metric. It is clear that even decomposition $\left(
\ref{dec}\right)  $ is affected by the reduction $\left(  \ref{Red}\right)  $
which induces a rearrangement of the Eq.$\left(  \ref{SepS2a}\right)  $. After
a lengthy algebraic manipulation we arrive at%
\begin{equation}
S^{\left(  2\right)  }=-\frac{1}{8\kappa}\int dt\int d^{3}xN\sqrt{g}\left[
\left(  h^{\bot}\right)  ^{ij}\left(  \tilde{\bigtriangleup}_{L\!}%
\!{}\,h^{\bot}\right)  _{ij}-\frac{2}{3}\sigma\tilde{\bigtriangleup}\!{}%
\sigma-\frac{2}{3}\sigma R_{jl}\left(  h^{\bot}\right)  ^{jl}\right]
,\label{S2r}%
\end{equation}
where%
\begin{equation}
\left(  \tilde{\bigtriangleup}_{L\!}\!{}h^{\bot}\right)  _{ij}=\left(
\bigtriangleup_{L\!}\!{}\,h^{\bot}\right)  _{ij}-4R{}_{i}^{k}\!{}h_{kj}^{\bot
}+R{}\!{}h_{ij}^{\bot}+\frac{1}{N^{2}}\frac{\partial^{2}}{\partial t^{2}%
}h_{ij}^{\bot}\label{M Lichn}%
\end{equation}
and%
\begin{equation}
\tilde{\bigtriangleup}\sigma=\bigtriangleup\!{}\sigma-\frac{1}{12}R{}\!{}%
\!{}\sigma-\frac{1}{2N^{2}}\frac{\partial^{2}}{\partial t^{2}}\sigma
.\label{M scalar}%
\end{equation}
It is immediate to recognize that for Einstein background $R_{ij}=Ag_{ij}$,
cross terms vanish. Unfortunately, the Schwarzschild metric in three
dimensions does not fall in this case. Nevertheless, the linearized action can
be represented in a short way on a suitable tensor space%
\begin{equation}
S^{\left(  2\right)  }=-\frac{1}{8\kappa}\int dt\int_{\Sigma}d^{3}xN\sqrt
{g}\left[  \left[  \left(  h^{\bot}\right)  ^{ij},\sigma\right]  O^{\left[
\left(  ij,m\right)  \left(  kl,n\right)  \right]  }\left[  \left(  h^{\bot
}\right)  ^{kl},\sigma\right]  ^{T}\right]  ,
\end{equation}
with $O^{\left[  \left(  ij,m\right)  \left(  kl,n\right)  \right]  }$ a
$\left[  \left(  3\times3\right)  +1\right]  \times\left[  \left(
3\times3\right)  +1\right]  $-matrix differential operator whose first
$\left(  3\times3\right)  $ block matrix act on transverse traceless spin two
field $h_{ij}^{\bot}$ and whose last columns acts on the spin zero field
$\sigma$. The corresponding matrix can be read off from $\left(
\ref{S2r}\right)  $%
\begin{equation}
O^{\left(  i,j\right)  }=%
\begin{bmatrix}
\tilde{\bigtriangleup}_{L\!} & -\frac{2}{3}R_{jl}\\
0 & \frac{2}{3}\tilde{\bigtriangleup}%
\end{bmatrix}
.
\end{equation}
To write the corresponding functional determinant, we observe that the
following relations are valid for arbitrary triangular matrix
operator\cite{KP}:%
\[
\ln\det\left(
\begin{array}
[c]{cc}%
A & C\\
0 & B
\end{array}
\right)  =Tr\ln\left(
\begin{array}
[c]{cc}%
A & C\\
0 & B
\end{array}
\right)  =Tr\left(
\begin{array}
[c]{cc}%
\ln A & C\\
0 & \ln B
\end{array}
\right)
\]%
\begin{equation}
=\ln A+\ln B=\ln\det\left(
\begin{array}
[c]{cc}%
A & 0\\
0 & B
\end{array}
\right)  .\label{lndet}%
\end{equation}
Thus, the mixing term does not come into play and we get%
\begin{equation}
\left(  \det\nolimits_{h^{\bot},\sigma}\left[  O^{\left(  i,j\right)
}\right]  \right)  ^{-\frac{1}{2}}=\left(  \det\nolimits_{\bot}\tilde
{\bigtriangleup}_{L\!}\right)  ^{-\frac{1}{2}}\left(  \det\nolimits_{\sigma
}\tilde{\bigtriangleup}\right)  ^{-\frac{1}{2}}.
\end{equation}
The same problem appears for the Jacobian. In Appendix \ref{AppLL}, we show
that Eq.$\left(  \ref{G1L}\right)  $ reduces to%
\begin{equation}
\Gamma_{1-loop}=\frac{i}{2}\int_{\mathcal{M}}d^{4}xN\sqrt{g}\int\frac{d^{4}%
k}{\left(  2\pi\right)  ^{4}}\ln\lambda_{TT}^{2}-\frac{i}{2}\int
_{\mathcal{\Sigma}}d^{3}x\sqrt{g}\int\frac{d^{3}k}{\left(  2\pi\right)  ^{3}%
}\ln\lambda_{V^{\bot}}^{2}%
\end{equation}
and Eq.$\left(  \ref{Tmn}\right)  $ changes into%
\begin{equation}
\left\langle T_{\mu\nu}^{C.C.}\right\rangle =-\frac{i}{2}g_{\mu\nu}\int
\frac{d^{4}k}{\left(  2\pi\right)  ^{4}}\ln\lambda_{TT}^{2}+\frac{i}{2}\left(
g_{\mu\nu}+u_{\mu}u_{\nu}\right)  \int\frac{d^{3}k}{\left(  2\pi\right)  ^{3}%
}\ln\lambda_{V^{\bot}}^{2},\label{LIndA}%
\end{equation}
where $u_{\mu}$ is a time-like unit vector. This means that the energy density
in 3+1 dimensions is not affected by the vector part which contributes only on
the pressure terms\footnote{This result is in agreement with the result of
Ref.\cite{GriKos}, where only the graviton contribution contributes to the
evaluation of the effective action.}. Eq.$\left(  \ref{LIndA}\right)  $.

\section{W.K.B. approximation of the functional determinants}

\label{p2}To evaluate $\rho_{GS}$ and $\rho_{W}$, we extract the energy
density from Eq.$\left(  \ref{LIndA}\right)  $ and we get%
\begin{equation}
\frac{\Lambda}{8\pi G}=-\frac{i}{2}\left[  \int\frac{d^{4}k}{\left(
2\pi\right)  ^{4}}\ln\lambda_{TT}^{2}\right]  =-\frac{i}{2}\sum_{i=1}^{2}%
\int\frac{d^{3}k}{\left(  2\pi\right)  ^{3}}\int_{-\infty}^{+\infty}%
\frac{d\omega}{2\pi}\ln\left(  -\frac{\omega_{i}^{2}}{N^{2}}+\lambda_{i}%
^{2}\left(  \left\vert \vec{k}\right\vert \right)  \right)  , \label{TrlnO}%
\end{equation}
where $\lambda_{i}^{2}\left(  \left\vert \vec{k}\right\vert \right)  $ are the
spatial eigenvalues of the operator $\tilde{\bigtriangleup}_{L}$. We use the
following formal representation to eliminate the logarithm\footnote{%
\begin{equation}
\ln\frac{b}{a}=\lim_{\varepsilon\rightarrow0}\int_{0}^{+\infty}\frac{dt}%
{t}e^{it\left(  a+i\varepsilon\right)  }-e^{it\left(  b+i\varepsilon\right)
},
\end{equation}
}%
\begin{equation}
\ln b=-\lim_{\varepsilon\rightarrow0}\int_{0}^{+\infty}\frac{dt}%
{t}e^{it\left(  b+i\varepsilon\right)  }.
\end{equation}
Eq.$\left(  \ref{TrlnO}\right)  $ can be cast into the form%
\[
\frac{i}{2}\int\frac{d^{3}k}{\left(  2\pi\right)  ^{3}}\int_{-\infty}%
^{+\infty}\frac{d\omega}{2\pi}\sum_{i=1}^{2}\lim_{\varepsilon\rightarrow0}%
\int_{0}^{+\infty}\frac{dt}{t}e^{-it\left(  \frac{\omega_{i}^{2}}{N^{2}%
}-\lambda_{i}^{2}\left(  \left\vert \vec{k}\right\vert \right)  +i\varepsilon
\right)  }=\frac{i}{2}\int\frac{d^{3}k}{\left(  2\pi\right)  ^{3}}\int
_{0}^{+\infty}\frac{dt}{t}\sum_{i=1}^{2}\lim_{\varepsilon\rightarrow0}%
\int_{-\infty}^{+\infty}\frac{d\omega}{2\pi}e^{-it\left(  \frac{\omega_{i}%
^{2}}{N^{2}}-\lambda_{i}^{2}\left(  \left\vert \vec{k}\right\vert \right)
+i\varepsilon\right)  }%
\]%
\begin{equation}
=\frac{iN}{4\sqrt{i\pi}}\int\frac{d^{3}k}{\left(  2\pi\right)  ^{3}}\sum
_{i=1}^{2}\lim_{\varepsilon\rightarrow0}\int_{0}^{+\infty}\frac{dt}%
{\sqrt{t^{3}}}e^{it\left(  \lambda_{i}^{2}\left(  \left\vert \vec
{k}\right\vert \right)  +i\varepsilon\right)  }=-\frac{N}{2}\sum_{i=1}^{2}%
\int\frac{d^{3}k}{\left(  2\pi\right)  ^{3}}\sqrt{\lambda_{i}^{2}\left(
\left\vert \vec{k}\right\vert \right)  }=\frac{\Lambda}{8\pi G},
\label{lambda}%
\end{equation}
where we have used the following representations%
\begin{equation}
\int_{-\infty}^{+\infty}\frac{d\omega}{2\pi}e^{-it\frac{\omega_{i}^{2}}{N^{2}%
}}=\frac{N}{2\sqrt{\pi it}}\qquad\text{and}\qquad\int_{0}^{+\infty}%
e^{ixt}t^{z-1}dt=\left(  -ix\right)  ^{-z}\Gamma\left(  z\right)  ,
\end{equation}
with $\operatorname{Im}\left(  x\right)  >0$ and $z=-\frac{1}{2}$. Note the
presence of the redshift function in Eq.$\left(  \ref{lambda}\right)  $. This
is a remnant of the original time component. In order to evaluate the integral
over momenta in Eq.$\left(  \ref{lambda}\right)  $, we use the WKB
approximation. With the help of Eqs.$\left(  \ref{potentials}\right)  $, we
define two r-dependent radial wave numbers $k_{1}\left(  r,l,\lambda
_{1,nl}\right)  $ and $k_{2}\left(  r,l,\lambda_{2,nl}\right)  $ for the
Lichnerowicz operator (TT tensor)%
\begin{equation}
\left\{
\begin{array}
[c]{c}%
k_{1}^{2}\left(  r,l,\lambda_{1,nl}\right)  =\lambda_{1,nl}^{2}-\frac{l\left(
l+1\right)  }{r^{2}}-m_{1}^{2}\left(  r\right) \\
\\
k_{2}^{2}\left(  r,l,\lambda_{2,nl}\right)  =\lambda_{2,nl}^{2}-\frac{l\left(
l+1\right)  }{r^{2}}-m_{2}^{2}\left(  r\right)
\end{array}
\right.  \label{kTT}%
\end{equation}
and we separate the effective masses in two pieces
\begin{equation}
\left\{
\begin{array}
[c]{c}%
m_{1}^{2}\left(  r\right)  =m_{L}^{2}\left(  r\right)  +m_{1,S}^{2}\left(
r\right) \\
\\
m_{2}^{2}\left(  r\right)  =m_{L}^{2}\left(  r\right)  +m_{2,S}^{2}\left(
r\right)
\end{array}
\right.  ,
\end{equation}
with%
\begin{equation}
m_{L}^{2}\left(  r\right)  =\frac{6}{r^{2}}\left(  1-\frac{b\left(  r\right)
}{r}\right)  \label{mL}%
\end{equation}
and%
\begin{equation}
\left\{
\begin{array}
[c]{c}%
m_{1,S}^{2}\left(  r\right)  =\frac{3}{2r^{2}}b^{\prime}\left(  r\right)
-\frac{3}{2r^{3}}b\left(  r\right) \\
m_{2,S}^{2}\left(  r\right)  =\frac{1}{2r^{2}}b^{\prime}\left(  r\right)
+\frac{3}{2r^{3}}b\left(  r\right)
\end{array}
\right.  . \label{ms}%
\end{equation}
The WKB approximation we will use is equivalent to the scattering phase shift
method and to the entropy computation in the brick wall model. We begin by
counting the number of modes with frequency less than $\lambda_{i}$, $i=1,2$.
This is given approximately by%
\begin{equation}
\tilde{g}\left(  \lambda_{i}\right)  =\int_{0}^{l_{\max}}\nu_{i}\left(
l,\lambda_{i}\right)  \left(  2l+1\right)  dl, \label{p41}%
\end{equation}
where $\nu_{i}\left(  l,\lambda_{i}\right)  $ is the number of nodes in the
mode with $\left(  l,\lambda_{i}\right)  $, such that $\left(  i=1,2\right)
$
\begin{equation}
\nu_{i}\left(  l,\lambda_{i}\right)  =\frac{1}{\pi}\int_{-\infty}^{+\infty
}dx\sqrt{k_{i}^{2}\left(  r,l,\lambda_{i}\right)  }.\qquad\left(  r\equiv
r\left(  x\right)  \right)  \label{p42}%
\end{equation}
In Eq.$\left(  \ref{p42}\right)  $ is understood that the integration with
respect to $x$ and $l$ is taken over those values which satisfy $k_{i}%
^{2}\left(  r,l,\lambda_{i}\right)  \geq0,$ $i=1,2$. With the help of
Eqs.$\left(  \ref{p41},\ref{p42}\right)  $, the total energy associated to the
energy density in Eq.$\left(  \ref{lambda}\right)  $ becomes $\left(  r\equiv
r\left(  x\right)  \right)  $%
\begin{equation}
N\sum_{i=1}^{2}\left[  \int_{0}^{+\infty}\lambda_{i}\frac{d\tilde{g}\left(
\lambda_{i}\right)  }{d\lambda_{i}}d\lambda_{i}\right]  =\frac{1}{4\pi^{2}%
}\sum_{i=1}^{2}\int_{-\infty}^{+\infty}dxN\left(  r\right)  r^{2}\int
_{\sqrt{m_{i}^{2}\left(  r\right)  }}^{+\infty}\lambda_{i}^{2}\sqrt
{\lambda_{i}^{2}-m_{i}^{2}\left(  r\right)  }d\lambda_{i}.
\end{equation}
By extracting the energy density, we obtain%
\begin{equation}
\frac{\Lambda}{8\pi G}=-\frac{1}{2}\sum_{i=1}^{2}\int\frac{d^{3}k}{\left(
2\pi\right)  ^{3}}\sqrt{\lambda_{i}^{2}\left(  \left\vert \vec{k}\right\vert
\right)  }=-\frac{1}{4\pi^{2}}\sum_{i=1}^{2}\int_{\sqrt{m_{i}^{2}\left(
r\right)  }}^{+\infty}\lambda_{i}^{2}\sqrt{\lambda_{i}^{2}-m_{i}^{2}\left(
r\right)  }d\lambda_{i}, \label{inte}%
\end{equation}
where we have included an additional $4\pi$ coming from the angular
integration and where we have included in the volume term the redshift
function. Of course, Eq.$\left(  \ref{inte}\right)  $ is divergent and must be regularized.

\section{Regularization and renormalization of one loop contribution to the
cosmological constant}

\label{p3}We adopt the zeta function regularization scheme and by introducing
the additional mass parameter $\mu$ in order to restore the correct dimension
for the regularized quantities, we define%
\begin{equation}
\rho_{i}\left(  \varepsilon,\mu\right)  =-\frac{1}{4\pi^{2}}\mu^{2\varepsilon
}\int_{\sqrt{m_{i}^{2}\left(  r\right)  }}^{+\infty}d\lambda_{i}\frac
{\lambda_{i}^{2}}{\left(  \lambda_{i}^{2}-m_{i}^{2}\left(  r\right)  \right)
^{\varepsilon-\frac{1}{2}}};\qquad i=1,2.\label{zeta}%
\end{equation}
The integration has to be meant in the range where $\lambda_{i}^{2}-m_{i}%
^{2}\left(  r\right)  \geq0$. Following the same steps as in Ref.\cite{Remo},
one gets%
\begin{equation}
\rho_{i}\left(  \varepsilon,\mu\right)  =\frac{m_{i}^{4}\left(  r\right)
}{64\pi^{2}}\left[  \frac{1}{\varepsilon}+\ln\left(  \frac{4\mu^{2}}{m_{i}%
^{2}\left(  r\right)  \sqrt{e}}\right)  \right]  ,\qquad i=1,2.
\end{equation}
In order to renormalize the divergent ZPE, we write%
\begin{equation}
\frac{\Lambda}{8\pi G}\rightarrow\frac{\Lambda_{0}}{8\pi G}+\frac
{\Lambda^{div}}{8\pi G}=\frac{\Lambda_{0}}{8\pi G}+\frac{m_{1}^{4}\left(
r\right)  +m_{i}^{4}\left(  r\right)  }{64\pi^{2}\varepsilon}.
\end{equation}
Thus, the renormalization is performed via the absorption of the divergent
part into the re-definition of the bare classical constant $\Lambda$. The
remaining finite value for the cosmological constant reads%
\begin{equation}
\frac{\Lambda_{0}\left(  \mu\right)  }{8\pi G}=\sum_{i=1}^{2}\rho_{i}\left(
\mu\right)  =\frac{1}{64\pi^{2}}\sum_{i=1}^{2}m_{i}^{4}\left(  r\right)
\ln\left(  \frac{4\mu^{2}}{m_{i}^{2}\left(  r\right)  \sqrt{e}}\right)
=\rho_{eff}^{TT}\left(  \mu,r\right)  .\label{lambda0}%
\end{equation}
To avoid the dependence on the arbitrary mass scale $\mu$ in Eq.$\left(
\ref{lambda0}\right)  $, we adopt the renormalization group equation and we
impose that\cite{RGeq}%
\begin{equation}
\frac{1}{8\pi G}\mu\frac{\partial\Lambda_{0}\left(  \mu\right)  }{\partial\mu
}=\mu\frac{d}{d\mu}\rho_{eff}^{TT}\left(  \mu,r\right)  .
\end{equation}
Solving it we find that the renormalized constant $\Lambda_{0}$ should be
treated as a running one in the sense that it varies provided that the scale
$\mu$ is changing
\begin{equation}
\frac{\Lambda_{0}\left(  \mu,r\right)  }{8\pi G}=\frac{\Lambda_{0}\left(
\mu_{0},r\right)  }{8\pi G}+\frac{m_{1}^{4}\left(  r\right)  +m_{2}^{4}\left(
r\right)  }{32\pi^{2}}\ln\frac{\mu}{\mu_{0}}.\label{RGsol}%
\end{equation}
Substituting Eq.$\left(  \ref{RGsol}\right)  $ into Eq.$\left(  \ref{lambda0}%
\right)  $ we find%
\begin{equation}
\frac{\Lambda_{0}\left(  \mu_{0},r\right)  }{8\pi G}=-\frac{1}{64\pi^{2}}%
\sum_{i=1}^{2}m_{i}^{4}\left(  r\right)  \ln\left(  \frac{m_{i}^{2}\left(
r\right)  \sqrt{e}}{4\mu_{0}^{2}}\right)  .\label{lambdamu0}%
\end{equation}
Potentially, we have three cases: $1)$ $m_{L}^{2}\left(  r\right)  \gg
m_{S}^{2}\left(  r\right)  $, $2)$ $m_{L}^{2}\left(  r\right)  =m_{S}%
^{2}\left(  r\right)  $ and $3)$ $m_{L}^{2}\left(  r\right)  \lll m_{S}%
^{2}\left(  r\right)  $. Case $2)$ reduces to a single point and therefore
will be discarded in this analysis. In case $1)$ essentially we consider a
long range contribution of the graviton which will be vanishing for
$r\rightarrow\infty$. Finally, case $3)$ is a short range case and it leads to%
\begin{equation}
\frac{\Lambda_{0}\left(  \mu_{0},r\right)  }{8\pi G}=-\frac{1}{64\pi^{2}}%
\sum_{i=1}^{2}\left[  m_{i,S}^{4}\left(  r\right)  \ln\left(  \frac
{m_{i,S}^{2}\left(  r\right)  }{4\mu_{0}^{2}}\sqrt{e}\right)  \right]
.\label{lambdaeff}%
\end{equation}
The above expression works for a background described by Eq.$\left(
\ref{metric}\right)  $ which must satisfy the Einstein's field equations. We
specialize the result to the case of interest, namely the Schwarzschild and
the de Sitter metrics. However, since the exterior part of the gravastar is of
the Schwarzschild form, we begin with this background which is in common with
a wormhole model. In the short range approximation we find%
\begin{equation}
m_{1,S}^{2}\left(  r\right)  =-m_{2,S}^{2}\left(  r\right)  =m_{S}^{2}\left(
r\right)  =\frac{3MG}{r^{3}}\label{masses}%
\end{equation}
and the range of validity is when $r\in\left[  2MG,5MG/2\right]  $ which can
be determined by case $2)$. In this range, $\Lambda_{0}\left(  \mu
_{0},r\right)  /\left(  8\pi G\right)  $ has the following properties:

\begin{description}
\item[i)] For $r=2MG$, $\Lambda_{0}\left(  \mu_{0},2MG\right)  \rightarrow
\infty$ when $M\rightarrow0$.

\item[ii)] For $r>2MG$, $\Lambda_{0}\left(  \mu_{0},r\right)  \rightarrow0$
when $M\rightarrow0$.
\end{description}

This can be summarized in the following double limit%
\begin{equation}
\lim_{M\rightarrow0}\lim_{r\rightarrow2MG}\Lambda_{0}\left(  \mu_{0},r\right)
\neq\lim_{r\rightarrow2MG}\lim_{M\rightarrow0}\Lambda_{0}\left(  \mu
_{0},r\right)  \label{NonComm}%
\end{equation}
which appears to be a sort of non-commutativity appearing in proximity of the
throat. A similar behavior was conjectured by Ahluwalia\cite{Ahluwalia} in
connection with the black hole entropy where a relation of the type%
\begin{equation}
\left[  l_{P,}l_{S}\right]  =i\lambda_{P}^{2}%
\end{equation}
was introduced. The \textquotedblleft Schwarzschild\textquotedblright\ $l_{S}$
and the \textquotedblleft Planck\textquotedblright\ $l_{P}$ lengths are no
more simply lengths, but operators. An analogy can also be found for
Yang-Mills theory in a constant chromomagnetic background\cite{NO}, even if
the situation interests the infrared region instead of the ultraviolet region.
This non-commutative effect of Eq.$\left(  \ref{NonComm}\right)  $ could be
interpreted as a signal of a phase transition.

\subsection{Persisting of the wormhole}

With the help of relations $\left(  \ref{masses}\right)  $ and from
Eq.$\left(  \ref{lambdaeff}\right)  $, setting $r=r_{t}=2MG$, we find that
Eq.$\left(  \ref{lambdaeff}\right)  $ becomes%
\begin{equation}
\frac{\Lambda_{0}\left(  \mu_{0},r_{t}\right)  }{8\pi G}=-\frac{9}{128\pi
^{2}r_{t}^{4}}\ln\left(  \frac{3\sqrt{e}}{8r_{t}^{2}\mu_{0}^{2}}\right)
,\label{lambda0t}%
\end{equation}
namely the throat does not manifest quantum fluctuations. Eq.$\left(
\ref{lambda0t}\right)  $ satisfies the following inequality%
\begin{equation}
\frac{\Lambda_{0}\left(  \mu_{0},r_{t}\right)  }{8\pi G}\leq\frac{\Lambda
_{0}\left(  \mu_{0},\bar{r}_{t}\right)  }{8\pi G}=\frac{9}{256\pi^{2}\bar
{r}_{t}^{4}}\qquad\mathrm{when\qquad}\sqrt{\frac{3e}{8\mu_{0}^{2}}}=\bar
{r}_{t},
\end{equation}
We recognize that the expression of the upper bound is of the Casimir form, in
the sense that we have a computation procedure mimicking the Casimir device
whose plates are located at the throat and at infinity. As in a Casimir device
whose energy density is proportional to the inverse fourth power distance of
the plates, also this case manifests the same behavior. From this it is
evident that the gravitational field of the wormhole can never be switched
off. However as $M$ becomes smaller and smaller, one cannot avoid to enter in
the quantum phase of the throat. Therefore, at a certain distance $r_{b}$ from
the throat, a \textquotedblleft\textit{brick wall}\textquotedblright%
\cite{tHooft}\ can be formed due to quantum fluctuations. This forbids the
throat to be reached. In some sense, it is the Casimir energy that changes the
structure of the throat. If this is the case, Eq.$\left(  \ref{lambda0t}%
\right)  $ becomes%
\begin{equation}
\frac{\Lambda_{0}\left(  \mu_{0},M,r_{b}\right)  }{8\pi G}=-\frac{1}{32\pi
^{2}}\left(  \frac{3MG}{r_{b}^{3}}\right)  ^{2}\ln\left(  \frac{3MG\sqrt{e}%
}{4r_{b}^{3}\mu_{0}^{2}}\right)  ,
\end{equation}
where $r_{b}$ is of the form $r_{b}=r_{t}+h$ with $h$ representing the
\textquotedblleft\textit{brick wall}\textquotedblright. The ZPE is now regular
when the throat vanishes.

\subsection{Wormhole turning to a gravastar}

In this case, the Casimir energy not only creates a \textquotedblleft%
\textit{brick wall}\textquotedblright, but changes completely the structure of
the wormhole by a topology change\footnote{See also Ref.\cite{DBGL} where a
discussion on the possible topology change induced by Casimir energy between
dark stars and wormholes is faced.}. The effect of the vacuum reorganization
induces a redefinition of the function $b\left(  r\right)  $ in $\left(
\ref{metric}\right)  $%
\begin{equation}
b\left(  r\right)  =\left\{
\begin{array}
[c]{cc}%
2MG\left(  \frac{r}{r_{0}}\right)  ^{3} & 0<r<r_{0}\\
2MG & r>r_{0}%
\end{array}
\right.  . \label{Newb(r)}%
\end{equation}
The \textquotedblleft\textit{inner cosmological constant}\textquotedblright%
\ is of course%
\begin{equation}
\Lambda_{inner}=\frac{6MG}{r_{0}^{3}}%
\end{equation}
and regulates the small de Sitter universe inside the wormhole. Of course the
new shape function is continuous in $r_{0}$, but it leads to a completely new
scenario, because it converts a wormhole throat in a \textquotedblleft%
\textit{cosmological throat}\textquotedblright, even if of very reduced size.
The metric is now regular at the origin and even if $b\left(  r\right)  $ is
continuous at $r_{0}$, the energy density is not. This is in agreement with
the initial setting of Eq.$\left(  \ref{GStar}\right)  $, but without the thin
shell. The same discontinuity reappears to one loop level for in the inner
region, Eq.$\left(  \ref{lambdaeff}\right)  $ becomes%
\begin{equation}
\frac{\Lambda_{0}\left(  \mu_{0},r_{0}\right)  _{|in}}{8\pi G}=-\frac{1}%
{32\pi^{2}}\left(  \frac{6MG}{r_{0}^{3}}\right)  ^{2}\ln\left(  \frac
{6MG\sqrt{e}}{4r_{0}^{3}\mu_{0}^{2}}\right)  , \label{in}%
\end{equation}
while in the outer region one gets%
\begin{equation}
\frac{\Lambda_{0}\left(  \mu_{0},r_{0}\right)  _{|out}}{8\pi G}=-\frac
{1}{32\pi^{2}}\left(  \frac{3MG}{r_{0}^{3}}\right)  ^{2}\ln\left(
\frac{3MG\sqrt{e}}{4r_{0}^{3}\mu_{0}^{2}}\right)  . \label{out}%
\end{equation}
However, if we try to insist and impose continuity on the boundary%
\begin{equation}
\Lambda_{0}\left(  \mu_{0},r_{0}\right)  _{|in}=\Lambda_{0}\left(  \mu
_{0},r_{0}\right)  _{|out}, \label{cont}%
\end{equation}
we find a solution provided one fixes the renormalization point to%
\begin{equation}
\frac{3MG\sqrt{e}}{\sqrt[3]{4}r_{0}^{3}}=\mu_{0}^{2}. \label{point}%
\end{equation}

\subsection{Comparing the gravastar to the wormhole}

We are now ready to compare the gravastar with the wormhole model. As
discussed in the introduction, we know that both models share the same
$\mathcal{ADM}$ mass. Essentially, the ADM energy is defined as%
\begin{equation}
E_{ADM}=\frac{1}{16\pi G}\int_{S}\left(  D^{i}h_{ij}-D_{j}h\right)  r^{j},
\label{ADM}%
\end{equation}
where the indices $i,j$ run over the three spatial dimensions and%
\begin{equation}
h_{ij}=g_{ij}-\bar{g}_{ij},
\end{equation}
where $\bar{g}_{ij}$ is the background three-metric. $D_{j}\ $is the
background covariant derivative and $r^{j}$ is the unit normal to the large
sphere $S$. However, Hawking and Horowitz\cite{HawHor} have shown that the
definition $\left(  \ref{ADM}\right)  $ is equivalent to%
\begin{equation}
E_{ADM}=\frac{1}{8\pi G}\int_{S^{\infty}}d^{2}x\sqrt{\sigma}\left(
k-k^{0}\right)  , \label{DeltaE}%
\end{equation}
where $\sigma$ is the determinant of the unit 2-sphere. $k^{0}$ represents the
trace of the extrinsic curvature corresponding to embedding in the
two-dimensional boundary $^{2}S$ in three-dimensional Euclidean space at
infinity. In Eq.$\left(  \ref{DeltaE}\right)  $, it is well represented the
subtraction procedure between two metrics having the same asymptotic behavior.
Therefore with a natural extension we define the subtraction procedure in such
a way that we can include quantum effects: this is the Casimir energy or in
other terms, the vacuum energy. One can in general formally define the Casimir
energy as follows%
\begin{equation}
E_{Casimir}\left[  \partial\mathcal{M}\right]  =E_{0}\left[  \partial
\mathcal{M}\right]  -E_{0}\left[  0\right]  , \label{a3}%
\end{equation}
where $E_{0}$ is the zero-point energy and $\partial\mathcal{M}$ is a
boundary. For zero temperature, the idea underlying the Casimir effect is to
compare vacuum energies in two physical distinct configurations. The extension
to quantum effects is straightforward
\begin{equation}
E_{Casimir}\left[  \partial\mathcal{M}\right]  =\left(  E_{0}\left[
\partial\mathcal{M}\right]  -E_{0}\left[  0\right]  \right)  _{classical}%
+\left(  E_{0}\left[  \partial\mathcal{M}\right]  -E_{0}\left[  0\right]
\right)  _{1-loop}+\ldots.
\end{equation}

In our picture, the classical part represented by the $\mathcal{ADM}$-like
energy is vanishing, because the asymptotic behavior is the same for both the
wormhole and the gravastar. This means that%
\begin{equation}
E_{Casimir}\left[  \partial\mathcal{M}\right]  =\left(  E_{0}\left[
\partial\mathcal{M}\right]  -E_{0}\left[  0\right]  \right)  _{1-loop}%
+\ldots.,
\end{equation}
namely $E_{Casimir}$ is governed by purely quantum fluctuations. Here with
$E_{Casimir}$ we mean that the energy is an energy density. Thus, the Casimir
energy can be regarded as a measure of the topology change, in the sense that
if $E_{Casimir}$ is positive then the topology change will be suppressed,
while if it is negative, it will be favored. It is important to remark that in
most physical situations, the Casimir energy is negative. Consider now the one
loop term and suppose to compare a gravastar and a wormhole with the same mass
$M$ and the same renormalization point $\mu_{0}$. If we take flat space as a
reference space, we can write%
\begin{equation}
\left(  E_{0}^{W}\left[  \partial\mathcal{M}\right]  -E_{0}^{GS}\left[
\partial\mathcal{M}\right]  \right)  _{1-loop}=\left(  E_{0}^{W}\left[
\partial\mathcal{M}\right]  -E_{0}\left[  0\right]  \right)  _{1-loop}+\left(
E_{0}\left[  0\right]  -E_{0}^{GS}\left[  \partial\mathcal{M}\right]  \right)
_{1-loop}\,, \label{Delta1}%
\end{equation}
where $E_{0}\left[  0\right]  _{1-loop}$ represents the ZPE contribution of
flat space to one loop, which is absorbed into the regularization procedure.
Since outside the gravastar radius $r_{0}$%
\begin{equation}
E_{0}^{W}\left[  \partial\mathcal{M}\right]  =E_{0}^{GS}\left[  \partial
\mathcal{M}\right]  ,
\end{equation}
we consider the region in proximity of $r_{0}$. There remains to be evaluated
the density energy difference%
\begin{equation}
\left(  E_{0}^{W}\left[  \partial\mathcal{M}\right]  -E_{0}^{GS}\left[
\partial\mathcal{M}\right]  \right)  _{1-loop}=-\frac{1}{8\pi G}\left(
\Lambda_{0}\left(  \mu_{0},r_{b}\right)  ^{W}-\Lambda_{0}\left(  \mu_{0}%
,r_{0}\right)  ^{GS}\right)  _{|in},
\end{equation}
where $\Lambda_{0}\left(  \mu_{0},r_{b}\right)  ^{W}$ means that we are
evaluating the one loop term due to the wormhole background in proximity of
the brick wall, where we expect to receive the largest energy density
contribution. The minus sign appears because of a consequence of the
definition of the induced cosmological constant of Eq.$\left(  \ref{effCC}%
\right)  $ and the definition of the energy density of Eq.$\left(
\ref{eff}\right)  $. We have to discuss when%
\begin{equation}
\left(  E_{0}^{W}\left[  \partial\mathcal{M}\right]  -E_{0}^{GS}\left[
\partial\mathcal{M}\right]  \right)  _{1-loop}\gtreqless0. \label{DeltaEWGS}%
\end{equation}
After algebraic manipulation, we find that this happens when%
\begin{equation}
x^{6}\ln\left(  \frac{x^{3}}{\sqrt[3]{16}}\right)  +\frac{4}{3}\ln2\ \left\{
\begin{array}
[c]{c}%
>0\\
=0\\
<0
\end{array}
\right.
\begin{array}
[c]{r}%
0<x<1;\qquad x>1.26\\
x=1;\qquad x=1.26\\
1<x<1.26
\end{array}
, \label{ine}%
\end{equation}
where we have used Eq.$\left(  \ref{point}\right)  $ and we have defined%
\begin{equation}
x=\frac{r_{0}}{r_{b}}. \label{ratio}%
\end{equation}

\begin{figure}[th]
\centering\includegraphics[width=3.0in]{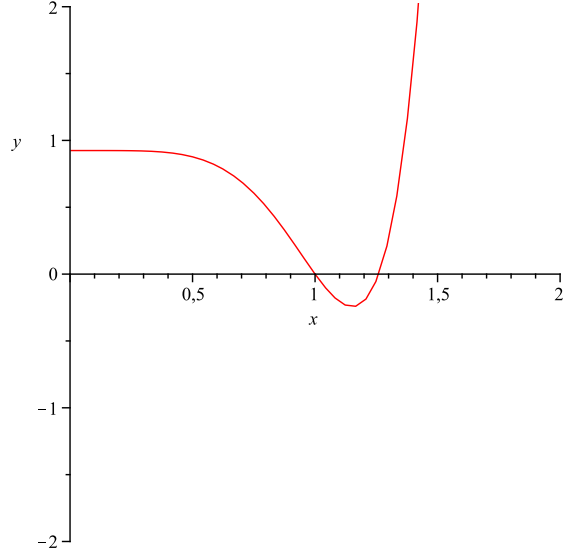}\caption{Plot of $\left(
E_{0}^{W}\left[  \partial\mathcal{M}\right]  -E_{0}^{GS}\left[  \partial
\mathcal{M}\right]  \right)  _{1-loop}$ as a function of $x=r_{0}/r_{b}.$}%
\label{Solfig}%
\end{figure}

The situation is better illustrated in Fig.$\left(  \ref{Solfig}\right)  $,
where we immediately recognize that a tiny region exists where the ZPE is
negative. This means that, in this range, the permanence of a wormhole is
energetically favored with respect to a gravastar of the same mass $M$. The
situation changes significantly if we avoid to fix the renormalization point
with the choice of Eq.$\left(  \ref{point}\right)  $ \begin{figure}[th]
\centering\includegraphics[width=3.0in]{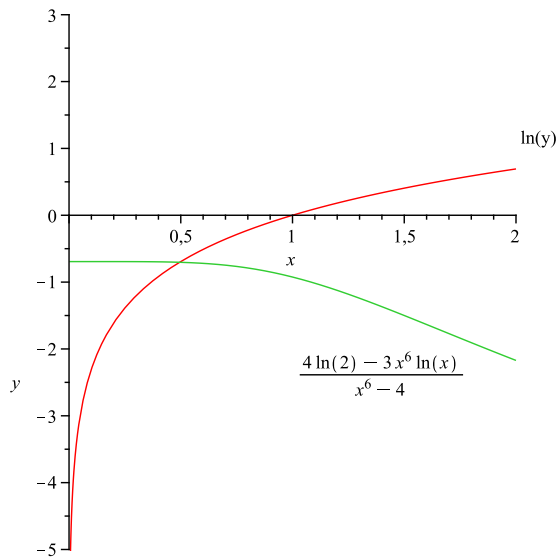}\caption{Plot of $\left(
E_{0}^{W}\left[  \partial\mathcal{M}\right]  -E_{0}^{GS}\left[  \partial
\mathcal{M}\right]  \right)  _{1-loop}$ as a function of $x=r_{0}/r_{b}$ and
$y$.}%
\label{sol1}%
\end{figure}This means that we are abandoning the continuity between the
external and the internal region of the gravastar to one loop. In this case,
Eq.$\left(  \ref{DeltaEWGS}\right)  $ becomes%
\begin{equation}
\ln y\ \gtreqless\frac{4\ln2-3x^{6}\ln x}{x^{6}-4},
\end{equation}
where $x$ is given again by Eq.$\left(  \ref{ratio}\right)  $ and%
\begin{equation}
y=\frac{3MG\sqrt{e}}{4r_{0}^{3}\mu_{0}^{2}}.
\end{equation}
The behavior of the ZPE difference is shown in Fig.$\left(  \ref{sol1}\right)
$, where the equality of expression $\left(  \ref{DeltaEWGS}\right)  $ is
reached when $\bar{y}=\bar{x}=0.4948$. It appears that below $\bar{y}$ and
$\bar{x}$, the ZPE becomes negative denoting that the \textquotedblleft%
\textit{ground state}\textquotedblright\ between a wormhole and a gravastar is
represented by a wormhole whose radius has to be smaller than the gravastar radius.

\section{Conclusions}

\label{p4}In this paper we have considered one loop corrections to the
$\mathcal{ADM}$ mass of a wormhole and a gravastar respectively. The
motivation comes from the fact that a gravastar can represent an alternative
to a black hole or a wormhole as regards the gravitational collapse.
Nevertheless various problems of comparison arise because outside the
gravastar radius, the metric is of the Schwarzschild type. If we adopt the
energy point of view we find that, asymptotically, these configurations share
the same $\mathcal{ADM}$ mass and therefore even in this case they are
indistinguishable. However, since the gravastar has a different core with
respect to the wormhole, a difference between them can emerge from a ZPE
contribution. To this purpose we have computed an effective action to one
loop, by looking at perturbations on the space-like hypersurface $\Sigma$.
Since the perturbation involves only the spatial part of the metric, ghosts do
not come into play for the energy contribution. Therefore, only the graviton
is important to establish what happens to the ZPE. Note that every form of ZPE
can be interpreted as an induced cosmological constant. In our case, this
interpretation is very useful to apply standard regularization and
renormalization procedures. In a sense, we can think that the ZPE induced by a
wormhole or a gravastar contributes to a cosmological constant. On the other
hand, we can think that the induced cosmological constant could be used to
give a sorting to ZPE. It is interesting to note that, it is the double limit
$\left(  \ref{NonComm}\right)  $ that denotes that something different appears
in the throat proximity. This difference is principally caused by ZPE or
Casimir energy which becomes so intense to create a thick barrier
(\textquotedblleft\textit{brick wall}\textquotedblright) or a topology change
(\textquotedblleft\textit{gravastar}\textquotedblright). Therefore, it is
important to discover under what condition a wormhole persists or change into
a gravastar or vice versa. It appears that a fundamental element to understand
in which direction the geometry becomes relevant is in the radii ratio
$\left(  \ref{ratio}\right)  $. Note that the comparison between the gravastar
and the wormhole is done with the same mass $M$, as it should be. The other
important parameter seems to be the renormalization point which, in this
example, translates the presence or the absence of an energy gap between the
external and the internal region of the gravastar. If one imposes the
continuity of the ZPE through the gravastar radius, one meets a constraint on
$\mu_{0}$ leading to the plot of Fig.$\left(  \ref{Solfig}\right)  $ showing
that the region of permanence of the wormhole is very subtle. Indeed, from
inequality $\left(  \ref{ine}\right)  $, outside the range $1<x<1.26$, the
wormhole appears as an \textquotedblleft\textit{excited state}%
\textquotedblright\ with respect to the gravastar. In particular, one should
note that the minimum for the ZPE difference appears for $x=1.152$ with a
minimum value of $-0.244$. On the other hand, if one abandons the continuity
condition of the gravastar ZPE and treats $\mu_{0}$ as a free parameter, one
finds from Fig.$\left(  \ref{sol1}\right)  $ that the region of permanence of
the wormhole is larger. Therefore we arrive at the conclusion that the
permanence of a wormhole or a gravastar in their reciprocal comparison is
strictly related to their size.

\section{Acknowledgement}

The author would like to thank R. Casadio, F.S.N. Lobo and D. Vassilevich for
useful comments, discussions and hints.

\appendix{}

\section{Disentangling the gauge modes}

\label{AppA}To explicitly make calculations, we need an orthogonal
decomposition for $h_{\mu\nu}$ to disentangle gauge modes from physical
deformations. To this purpose it is convenient to decompose $h_{\mu\nu}$ into
a trace, longitudinal and transverse-traceless part in $D$
dimensions\cite{MM1,Vassilevich,Quad}:%

\begin{equation}
h_{\mu\nu}=\frac{1}{D}hg_{\mu\nu}+\left(  L\xi\right)  _{\mu\nu}+h_{\mu\nu
}^{\bot}, \label{dec}%
\end{equation}
where the operator $L$ maps $\xi_{\mu}$ into symmetric tracefree tensors%

\begin{equation}
\left(  L\xi\right)  _{\mu\nu}=\nabla_{\mu}\xi_{\nu}+\nabla_{\nu}\xi_{\mu
}-\frac{2}{D}g_{\mu\nu}\left(  \nabla\cdot\xi\right)  \label{gaugev}%
\end{equation}
and%
\begin{equation}
g^{\mu\nu}h_{\mu\nu}^{\bot}=0,\qquad\nabla^{\mu}h_{\mu\nu}^{\bot}=0.
\end{equation}
The decomposition $\left(  \ref{dec}\right)  $ is orthogonal with respect to
the following inner product%
\begin{equation}
\left\langle h,k\right\rangle :=\int_{\mathcal{M}}d^{D}x\sqrt{-g}G^{\mu\nu
\rho\sigma}h_{\mu\nu}\left(  x\right)  k_{\rho\sigma}\left(  x\right)  ,
\label{inpr4d}%
\end{equation}
where%

\begin{equation}
G^{\mu\nu\rho\sigma}=\frac{1}{2}(g^{\mu\rho}g^{\nu\sigma}+g^{\mu\sigma}%
g^{\nu\rho}+Cg^{\mu\nu}g^{\rho\sigma}) \label{superm}%
\end{equation}
and $C$ is a constant. For the positivity of $G^{\mu\nu\rho\sigma}$ $C>-1/2$.
The inverse metric is defined on cotangent space and it assumes the form%

\begin{equation}
\left\langle p,q\right\rangle :=\int_{\mathcal{M}}d^{D}x\sqrt{-g}G_{\mu\nu
\rho\sigma}p^{\mu\nu}\left(  x\right)  q^{\rho\sigma}\left(  x\right)  ,
\end{equation}
so that%

\begin{equation}
G^{\mu\nu\alpha\beta}G_{\alpha\beta\rho\sigma}=\frac{1}{2}\left(  \delta
_{\rho}^{\mu}\delta_{\sigma}^{\nu}+\delta_{\sigma}^{\mu}\delta_{\rho}^{\nu
}\right)  .
\end{equation}
Following Ref.\cite{Vassilevich}, we observe that under the action of
infinitesimal diffeomorphism generated by a vector field $\epsilon_{\mu}$, the
components of $\left(  \ref{dec}\right)  $ transform as follows%
\begin{equation}
\xi_{\mu}\longrightarrow\xi_{\mu}+\epsilon_{\mu},\qquad h\longrightarrow
h+2\nabla^{\mu}\epsilon_{\mu},\qquad h_{\mu\nu}^{\bot}\longrightarrow
h_{\mu\nu}^{\bot}. \label{gauge}%
\end{equation}
We can fix the gauge freedom $\left(  \ref{gauge}\right)  $, by fixing%
\begin{equation}
\xi_{\mu}=0.
\end{equation}
If the manifold admits conformal Killing vectors, namely vectors annihilated
by the operator $L$, one additional gauge condition involving the trace is
necessary. Assume that such vectors are absent form the manifold. The Jacobian
factor induced by the change of variable, namely $h_{\mu\nu}\rightarrow\left(
h,\xi_{\mu},h_{\mu\nu}^{\bot}\right)  $ is%
\begin{equation}
J=\det_{V}\left(  L^{\dagger}L\right)  ^{\frac{1}{2}},
\end{equation}
where the determinant is calculated in the space of vector fields excluding
conformal Killing vectors. Using the definition $\left(  \ref{Lapl}\right)  $,
the operator acting on vector fields is
\begin{equation}
\left(  L^{\dagger}L\right)  _{\mu}^{\nu}=-2\left(  -\bigtriangleup\delta
_{\mu}^{\nu}+\left(  1-\frac{2}{D}\right)  \nabla_{\mu}\nabla^{\nu}+R_{\mu
}^{\nu}\right)  . \label{LLVec}%
\end{equation}
We can write decomposition $\left(  \ref{dec}\right)  $ in the following way%
\begin{equation}
h_{\mu\nu}=\left(  \sigma+2\nabla\cdot\xi\right)  \frac{g_{\mu\nu}}{D}+\left(
L\xi\right)  _{\mu\nu}+h_{\mu\nu}^{\bot},
\end{equation}
The change of variables $h\rightarrow\sigma$ does not introduce any additional
Jacobian factor, then the path integral measure separates into%
\begin{equation}
\mathcal{D}h_{\mu\nu}=\det_{V}\left(  L^{\dagger}L\right)  ^{\frac{1}{2}%
}\mathcal{D}h^{\bot}\mathcal{D}\xi\mathcal{D}\sigma
\end{equation}
and the quadratic part of the action $\left(  \ref{S24d}\right)  $ becomes%
\begin{equation}
S^{\left(  2\right)  }=\frac{1}{2\kappa}\int_{\mathcal{M}}d^{4}x\sqrt
{-g}\left[  -\frac{1}{4}h^{\bot,\mu\nu}\left(  \bigtriangleup_{L\!}%
\!{}h\right)  _{\mu\nu}^{\bot}+\frac{3}{32}\sigma\bigtriangleup\!{}%
\sigma\right]  . \label{S24da}%
\end{equation}
The path integral can be written in terms of functional determinants of
transverse-traceless tensor fields (indicated by T), vector fields (indicated
by V) and scalar fields (indicated by S):%
\begin{equation}
Z=\det_{T}\left(  \bigtriangleup_{L\!}\right)  ^{-\frac{1}{2}}\det_{V}\left(
L^{\dagger}L\right)  ^{\frac{1}{2}}\det_{S}\left(  -\bigtriangleup\right)
^{-\frac{1}{2}}.
\end{equation}
The determinant over vector fields is the analogue of the Faddeev-Popov
determinant. We can further decompose it by introducing a Hodge decomposition%
\begin{equation}
\xi=d\psi+\xi^{H}+\delta\omega=\xi^{||}+\xi^{H}+\xi^{\bot},
\end{equation}
where $\psi$ is a zero-form, $\xi^{H}$ is a harmonic one-form and $\omega$ is
a two-form. Excluding the presence of harmonic vectors $\xi^{H}$, we find that
the operator $\left(  \ref{LLVec}\right)  $ separates into%
\begin{equation}
\left(  L^{\dagger}L\xi\right)  _{\mu}=2\left(  \bigtriangleup\xi_{\mu}^{\bot
}-R_{\mu}^{\nu}\xi_{\mu}^{\bot}\right)  +2\left(  \bigtriangleup\xi_{\mu}%
^{||}-\left(  1-\frac{2}{D}\right)  \nabla_{\mu}\nabla^{\nu}\xi_{\nu}%
^{||}-R_{\mu}^{\nu}\xi_{\nu}^{||}\right)  \label{LL4D}%
\end{equation}
and using the equations of motion on shell $R_{\mu\nu}=0$, we get for the
Jacobian%
\begin{equation}
\det_{V}\left(  L^{\dagger}L\right)  ^{\frac{1}{2}}=\det_{V^{\bot}}\left(
\bigtriangleup\right)  ^{\frac{1}{2}}\det_{V^{||}}\left(  \bigtriangleup
\right)  ^{\frac{1}{2}} \label{dLL4D}%
\end{equation}
and the one loop effective action simplifies
\begin{equation}
\Gamma_{1-loop}=\frac{i}{2}\left[  Tr\ln\bigtriangleup_{L\!}^{\bot}%
\!-Tr\ln\triangle_{V^{\bot}}\right]  =\frac{i}{2}\int_{\mathcal{M}}d^{4}%
x\sqrt{-g}\left[  \int\frac{d^{4}k}{\left(  2\pi\right)  ^{4}}\ln\lambda
_{TT}^{2}-\int\frac{d^{4}k}{\left(  2\pi\right)  ^{4}}\ln\lambda_{V^{\bot}%
}^{2}\right]  ,
\end{equation}
where $\lambda_{TT}^{2}$ and $\lambda_{V^{\bot}}^{2}$ are the eigenvalues of
the Lichnerowicz operator for TT tensors and the transverse vector operator
respectively. Before going on, we need to precise a point concerning the
computation of a functional determinant. Generally speaking the functional
determinant of a given differential operator $O$ can be represented by%
\begin{equation}
\det O=\exp\left(  Tr\ln O\right)  .
\end{equation}
However within the zeta function regularization, it is no longer true that%
\begin{equation}
\det AB=\det A\det B,
\end{equation}
where $A$ and $B$ are two elliptic operators. In general, one has%
\begin{equation}
\det AB=\exp a\left(  A,B\right)  \det A\det B.
\end{equation}
where $a\left(  A,B\right)  $ is a local functional called multiplicative
anomaly. As pointed out in Ref.\cite{EVZ}, one can assume the multiplicative
anomaly to be trivial, namely $a\left(  A,B\right)  =0$. This is justified by
the fact that to one loop approximation a non-trivial multiplicative anomaly
may be absorbed into the renormalization ambiguity.

\subsection{Evaluating $\det_{V}\left(  L^{\dagger}L\right)  $ in $3$
dimensions}

\label{AppLL}In Eqs.$\left(  \ref{dec}\right)  $, $\left(  \ref{gaugev}%
\right)  $ and $\left(  \ref{LL4D}\right)  $, we have simply to put $D=3$. In
addition, we have to remind that, in this case the Ricci tensor is not
vanishing. If $\xi_{a}$ is further decomposed into a transverse part $\xi
_{a}^{T}$ with $\nabla^{a}\xi_{a}^{T}=0$ and a longitudinal part $\xi
_{a}^{\parallel}$ with $\xi_{a}^{\parallel}=$ $\nabla_{a}\psi$, then the
orthogonal decomposition reduces to%
\begin{equation}
\left\langle L\xi,L\xi\right\rangle :=\int_{\mathcal{M}}d^{3}x\sqrt{-g}\left(
L\xi\right)  ^{ij}\left(  L\xi\right)  _{ij} \label{p21b}%
\end{equation}%
\begin{equation}
=\int_{\mathcal{M}}d^{D}x\sqrt{-g}\left[  2\xi_{i}^{T}\bigtriangleup_{V}%
^{ij}\xi_{j}^{T}-4\xi_{i}^{T}R^{ij}\nabla_{j}\psi-4\psi\left(  \frac{2}%
{3}\bigtriangleup^{2}+\nabla_{i}R^{ij}\nabla_{j}\right)  \psi\right]  ,
\label{hh}%
\end{equation}
with%
\begin{equation}
\bigtriangleup_{V}^{ij}=\bigtriangleup g^{ij}-R^{ij}.
\end{equation}
The decomposition $\left(  \ref{hh}\right)  $ is orthogonal up to the $\xi
_{j}^{T}-\psi$-mixed terms. If we try to compute the related Jacobian induced
by the vector-scalar part in Eq.$\left(  \ref{hh}\right)  $, we obtain%
\[
\int\mathcal{D}\xi\exp\left[  -\frac{i}{2}\left\langle L\xi,L\xi\right\rangle
\right]  =
\]%
\begin{equation}
J_{1}\int\mathcal{D}\xi\mathcal{D}\psi\exp\left[  -\frac{i}{2}\int
_{\mathcal{M}}d^{D}x\sqrt{-g}\left\{  \left[  \xi_{i}^{T},\psi\right]
M^{\left(  i,j\right)  }\left[  \xi_{j}^{T},\psi\right]  ^{T}\right\}
\right]  =1.
\end{equation}
In terms of the functional determinant%
\begin{equation}
J_{1}=\left(  \det\nolimits_{\xi^{T},\psi}\left[  M^{\left(  i,j\right)
}\right]  \right)  ^{\frac{1}{2}},
\end{equation}
where $M^{\left(  i,j\right)  }$ a $\left(  3+1\right)  \times\left(
3+1\right)  $-matrix differential operator whose first 3 columns act on
transverse spin one field $\xi_{j}^{T}$ and whose last columns acts on the
spin zero field $\psi$. From Eq.$\left(  \ref{hh}\right)  $, the corresponding
matrix can be read off,%
\begin{equation}
M^{\left(  i,j\right)  }=%
\begin{bmatrix}
2\bigtriangleup_{V}^{T,ij} & -4R^{ij}\nabla_{j}\\
0 & -2\left(  \frac{2}{3}\bigtriangleup^{2}+\nabla_{i}R^{ij}\nabla_{j}\right)
\end{bmatrix}
. \label{Mij}%
\end{equation}
Thus Eq.$\left(  \ref{dLL4D}\right)  $ becomes%
\begin{equation}
\left(  \det\nolimits_{\xi^{T},\psi}\left[  M^{\left(  i,j\right)  }\right]
\right)  ^{\frac{1}{2}}=\left(  \det\bigtriangleup_{V}^{T}\right)  ^{\frac
{1}{2}}\left(  \det\left[  \frac{2}{3}\bigtriangleup^{2}+\nabla_{i}%
R^{ij}\nabla_{j}\right]  \right)  ^{\frac{1}{2}}. \label{detM}%
\end{equation}
The presence of the Ricci tensor in the scalar term of Eq.$\left(
\ref{detM}\right)  $ is an artifact of the foliation, because in 4 dimensions
it disappears. By means of the contracted Bianchi identities $\nabla_{i}%
R^{ij}=\nabla^{j}R/2$, we can simplify somewhat the expression of the scalar
term of Eq.$\left(  \ref{detM}\right)  $%
\begin{equation}
\frac{2}{3}\bigtriangleup^{2}+\nabla_{i}R^{ij}\nabla_{j}=\frac{2}%
{3}\bigtriangleup^{2}+\frac{1}{2}\nabla^{j}R\nabla_{j}+R^{ij}\nabla_{i}%
\nabla_{j}. \label{JS}%
\end{equation}
The determinant of the operator in Eq.$\left(  \ref{JS}\right)  $ can be cast
into the following form%
\[
\det\left(  \frac{2}{3}\bigtriangleup^{2}+R^{ij}\nabla_{i}\nabla_{j}\right)
^{\frac{1}{2}}=\exp\frac{1}{2}Tr\ln\left(  \frac{2}{3}\bigtriangleup
^{2}+R^{ij}\nabla_{i}\nabla_{j}\right)
\]%
\[
=\exp\frac{1}{2}Tr\ln\left[  \left(  \frac{2}{3}\bigtriangleup^{2}\right)
\left(  \mathbf{1}\mathbb{+}\left(  \frac{2}{3}\bigtriangleup^{2}\right)
^{-1}+R^{ij}\nabla_{i}\nabla_{j}\right)  \right]
\]%
\begin{equation}
=\exp\frac{1}{2}Tr\left[  \ln\left(  \frac{2}{3}\bigtriangleup^{2}\right)
+\ln\left(  \mathbf{1}\mathbb{+}\left(  \frac{2}{3}\bigtriangleup^{2}\right)
^{-1}R^{ij}\nabla_{i}\nabla_{j}\right)  \right]  .
\end{equation}
The second term is a correction to the principal part and will not be
considered in the W.K.B. method. The main part of the operator reduces to%
\begin{equation}
\det\left(  \frac{2}{3}\bigtriangleup^{2}\right)  ^{\frac{1}{2}}=\left(
\det\bigtriangleup\right)  ^{\frac{1}{2}}, \label{JS1}%
\end{equation}
where we have absorbed the constant factor into the definition of the
determinant and we have redefined the scalar wave function in such a way to
absorb the operator $\bigtriangleup$. To summarize, Eq.$\left(  \ref{dLL4D}%
\right)  $ in 3 dimensions becomes%
\begin{equation}
\det_{V}\left(  L^{\dagger}L\right)  ^{\frac{1}{2}}=\left(  \det
\bigtriangleup_{V}^{T}\right)  ^{\frac{1}{2}}\left(  \det\bigtriangleup
\right)  ^{\frac{1}{2}}. \label{dLL3D}%
\end{equation}
With the help of the functional determinant in Eq.$\left(  \ref{dLL3D}\right)
$, the one loop effective action can be written as%
\begin{equation}
\Gamma_{1-loop}=\left(  \det\bigtriangleup_{V}^{T}\right)  ^{\frac{1}{2}}%
\det\left(  \bigtriangleup\right)  ^{\frac{1}{2}}\left(  \det\tilde
{\bigtriangleup}_{L\!}\right)  ^{-\frac{1}{2}}\left(  \det\tilde
{\bigtriangleup}_{\sigma}\right)  ^{-\frac{1}{2}}. \label{Z2}%
\end{equation}
One can observe that the scalar determinants should cancel each other except
for three subtleties. First the sign of the operator $\tilde{\bigtriangleup
}\sigma$ in $\left(  \ref{M scalar}\right)  $ appears to be different from
that in the Jacobian, second the integration leading to $\left(
\ref{detM}\right)  $ excludes zero modes not included in the Jacobian, so any
cancellation will not be complete. This also happens in the full covariant
computation. Last but not least, the cancellation should be done after
integration over the time part in the determinant of the operator.

\section{The Lichnerowicz operator for TT tensors}

\label{appT}Our starting point is the expression $\left(  \ref{quadr}\right)
$ and the metric $\left(  \ref{metric}\right)  $. For the benefit of the
reader, we recall the representation of the operator $O$%
\begin{equation}
O^{ikjl}=\bigtriangleup_{L}^{ikjl}-4R^{il}g^{kj}+Rg^{ik}g^{jl}+\frac
{\partial^{2}}{\partial t^{2}}g^{ik}g^{jl}%
\end{equation}
and we simplify the expression of the Riemann tensor in $3$ dimensions%
\begin{equation}
R_{ikjl}=g_{ij}R_{kl}-g_{il}R_{kj}-g_{kj}R_{il}+g_{kl}R_{ij}-\frac{R}%
{2}\left(  g_{ij}g_{kl}-g_{il}g_{kj}\right)  .
\end{equation}
Then, the operator $O^{ikjl}$ becomes%
\[
-\nabla^{a}\nabla_{a}g^{ik}g^{jl}+\frac{\partial^{2}}{\partial t^{2}}%
g^{ik}g^{jl}-2\left(  g^{ij}R^{kl}-g^{il}R^{kj}-g^{kj}R^{il}+g^{kl}%
R^{ij}\right)  +R\left(  g^{ij}g^{kl}-g^{il}g^{kj}\right)
\]%
\[
+R^{ik}g^{jl}+R^{jk}g^{il}-4R^{il}g^{kj}+Rg^{ik}g^{jl}=-\nabla^{a}\nabla
_{a}g^{ik}g^{jl}+\frac{\partial^{2}}{\partial t^{2}}g^{ik}g^{jl}%
\]%
\begin{equation}
-2\left(  g^{ij}R^{kl}+g^{kl}R^{ij}\right)  +R\left(  g^{ij}g^{kl}%
-g^{il}g^{kj}\right)  +R^{ik}g^{jl}+3R^{jk}g^{il}-2R^{il}g^{kj}+Rg^{ik}g^{jl}.
\end{equation}
When we fix our attention on TT tensors, we obtain a further reduction%
\begin{equation}
O^{ikjl}=-\nabla^{a}\nabla_{a}g^{ik}g^{jl}+\frac{1}{N^{2}}\frac{\partial^{2}%
}{\partial t^{2}}g^{ik}g^{jl}+2R^{il}g^{kj}.
\end{equation}
Thus the related eigenvalue equation is%
\begin{equation}
-\left(  \triangle_{2}h^{TT}\right)  _{i}^{j}+\frac{1}{N^{2}}\frac
{\partial^{2}}{\partial t^{2}}h_{i}^{j}=\lambda^{2}h_{i}^{j} \label{p31}%
\end{equation}
where%
\begin{equation}
\left(  \triangle_{2}h^{TT}\right)  _{i}^{j}:=\left(  -\triangle_{T}%
h^{TT}\right)  _{i}^{j}+2\left(  Rh^{TT}\right)  _{i}^{j}%
\end{equation}
and%
\begin{equation}
-\left(  \triangle_{T}h^{TT}\right)  _{i}^{j}=-\triangle_{S}\left(
h^{TT}\right)  _{i}^{j}+\frac{6}{r^{2}}\left(  1-\frac{b\left(  r\right)  }%
{r}\right)  \left(  h^{TT}\right)  _{i}^{j}. \label{tlap}%
\end{equation}
$\triangle_{S}$ is the scalar curved Laplacian computed on the background of
metric $\left(  \ref{metric}\right)  $, whose form is%
\begin{equation}
\triangle_{S}=\left(  1-\frac{b\left(  r\right)  }{r}\right)  \frac{d^{2}%
}{dr^{2}}+\left(  \frac{4r-b^{\prime}\left(  r\right)  r-3b\left(  r\right)
}{2r^{2}}\right)  \frac{d}{dr}-\frac{L^{2}}{r^{2}} \label{deltaS}%
\end{equation}
and $R_{j\text{ }}^{a}$ is the mixed Ricci tensor whose components are:
\begin{equation}
R_{i}^{a}=\left\{  \frac{b^{\prime}\left(  r\right)  }{r^{2}}-\frac{b\left(
r\right)  }{r^{3}},\frac{b^{\prime}\left(  r\right)  }{2r^{2}}+\frac{b\left(
r\right)  }{2r^{3}},\frac{b^{\prime}\left(  r\right)  }{2r^{2}}+\frac{b\left(
r\right)  }{2r^{3}}\right\}  . \label{Ricci}%
\end{equation}
We will follow Regge and Wheeler in analyzing the equation as modes of
definite frequency, angular momentum and parity\cite{RW}. In particular, our
choice for the three-dimensional gravitational perturbation is represented by
its even-parity form%

\begin{equation}
h_{ij}^{even}\left(  t,r,\vartheta,\phi\right)  =diag\left[  H\left(
r\right)  \left(  1-\frac{b\left(  r\right)  }{r}\right)  ^{-1},r^{2}K\left(
r\right)  ,r^{2}\sin^{2}\vartheta K\left(  r\right)  \right]  Y_{lm}\left(
\vartheta,\phi\right)  F\left(  t\right)  . \label{pert}%
\end{equation}
For a generic value of the angular momentum $L$ representation $\left(
\ref{pert}\right)  $, together with Eqs.$\left(  \ref{tlap},\ref{p31}\right)
$ leads to the following system of PDE's%

\begin{equation}
\left\{
\begin{array}
[c]{c}%
\left(  -\triangle_{l}+2\left(  \frac{b^{\prime}\left(  r\right)  }{r^{2}%
}-\frac{b\left(  r\right)  }{r^{3}}\right)  +\frac{\partial^{2}}{\partial
t^{2}}\right)  H\left(  r\right)  F\left(  t\right)  =\lambda_{1,l}%
^{2}H\left(  r\right)  F\left(  t\right) \\
\\
\left(  -\triangle_{l}+2\left(  \frac{b^{\prime}\left(  r\right)  }{2r^{2}%
}+\frac{b\left(  r\right)  }{2r^{3}}\right)  +\frac{\partial^{2}}{\partial
t^{2}}\right)  K\left(  r\right)  F\left(  t\right)  =\lambda_{2,l}%
^{2}K\left(  r\right)  F\left(  t\right)
\end{array}
\right.  , \label{p33}%
\end{equation}
where $\triangle_{l}$ is%

\begin{equation}
\triangle_{l}=\left(  1-\frac{b\left(  r\right)  }{r}\right)  \frac{d^{2}%
}{dr^{2}}+\left(  \frac{4r-b^{\prime}\left(  r\right)  r-3b\left(  r\right)
}{2r^{2}}\right)  \frac{d}{dr}-\frac{l\left(  l+1\right)  }{r^{2}}-\frac
{6}{r^{2}}\left(  1-\frac{b\left(  r\right)  }{r}\right)  . \label{p33a}%
\end{equation}
The action of $\triangle_{l}$ on the reduced fields%

\begin{equation}
H\left(  r\right)  =\frac{f_{1}\left(  r\right)  }{r};\qquad K\left(
r\right)  =\frac{f_{2}\left(  r\right)  }{r}%
\end{equation}
is%
\[
\triangle_{l}\left(  \frac{f_{1,2}\left(  r\right)  }{r}\right)  =\frac{1}%
{r}\left\{  \left(  1-\frac{b\left(  r\right)  }{r}\right)  \frac{d^{2}%
}{dr^{2}}+\frac{1}{r}\left[  -2\left(  1-\frac{b\left(  r\right)  }{r}\right)
+\left(  \frac{4r-b^{\prime}\left(  r\right)  r-3b\left(  r\right)  }%
{2r}\right)  \right]  \frac{d}{dr}\right.
\]%
\[
\left.  +\frac{1}{r^{2}}\left[  2\left(  1-\frac{b\left(  r\right)  }%
{r}\right)  -\left(  \frac{4r-b^{\prime}\left(  r\right)  r-3b\left(
r\right)  }{2r}\right)  \right]  -\frac{l\left(  l+1\right)  }{r^{2}}-\frac
{6}{r^{2}}\left(  1-\frac{b\left(  r\right)  }{r}\right)  \right\}
f_{1,2}\left(  r\right)
\]%
\begin{equation}
=\frac{1}{r}\left\{  \left(  1-\frac{b\left(  r\right)  }{r}\right)
\frac{d^{2}}{dr^{2}}+\frac{1}{r}\left[  -\frac{b^{\prime}\left(  r\right)
}{2}+\frac{b\left(  r\right)  }{2r}\right]  \frac{d}{dr}+\frac{1}{r^{2}%
}\left[  \frac{b^{\prime}\left(  r\right)  }{2}-\frac{b\left(  r\right)  }%
{2r}\right]  -\frac{l\left(  l+1\right)  }{r^{2}}-\frac{6}{r^{2}}\left(
1-\frac{b\left(  r\right)  }{r}\right)  \right\}  f_{1,2}\left(  r\right)
\end{equation}
and using the proper geodesic distance from the \textit{throat}%
\begin{equation}
dx=\pm\frac{dr}{\sqrt{1-\frac{b\left(  r\right)  }{r}}}, \label{dt}%
\end{equation}
we get%
\begin{equation}
=\frac{1}{r}\left\{  \frac{d^{2}}{dx^{2}}+\frac{1}{r^{2}}\left[
\frac{b^{\prime}\left(  r\right)  }{2}-\frac{b\left(  r\right)  }{2r}\right]
-\frac{l\left(  l+1\right)  }{r^{2}}-\frac{6}{r^{2}}\left(  1-\frac{b\left(
r\right)  }{r}\right)  \right\}  f_{1,2}\left(  r\right)  ,
\end{equation}
where $r\equiv r\left(  x\right)  $. Thus, the system $\left(  \ref{p33}%
\right)  $ becomes%

\begin{equation}
\left\{
\begin{array}
[c]{c}%
\left[  -\frac{d^{2}}{dx^{2}}+\frac{\partial^{2}}{\partial t^{2}}+V_{1}\left(
r\right)  \right]  f_{1}\left(  x\right)  F\left(  t\right)  =\lambda
_{1,l}^{2}f_{1}\left(  x\right)  F\left(  t\right) \\
\\
\left[  -\frac{d^{2}}{dx^{2}}+\frac{\partial^{2}}{\partial t^{2}}+V_{2}\left(
r\right)  \right]  f_{2}\left(  x\right)  F\left(  t\right)  =\lambda
_{2,l}^{2}f_{2}\left(  x\right)  F\left(  t\right)
\end{array}
\right.  \label{F p33}%
\end{equation}
with
\begin{equation}
\left\{
\begin{array}
[c]{c}%
V_{1}\left(  r\right)  =\frac{l\left(  l+1\right)  }{r^{2}}+U_{1}\left(
r\right) \\
\\
V_{2}\left(  r\right)  =\frac{l\left(  l+1\right)  }{r^{2}}+U_{2}\left(
r\right)
\end{array}
\right.  ,
\end{equation}
and%
\begin{equation}
\left\{
\begin{array}
[c]{c}%
U_{1}\left(  r\right)  =\frac{6}{r^{2}}\left(  1-\frac{b\left(  r\right)  }%
{r}\right)  +\left[  \frac{3}{2r^{2}}b^{\prime}\left(  r\right)  -\frac
{3}{2r^{3}}b\left(  r\right)  \right]  =m_{1}^{2}\left(  r\right) \\
U_{2}\left(  r\right)  =\frac{6}{r^{2}}\left(  1-\frac{b\left(  r\right)  }%
{r}\right)  +\left[  \frac{1}{2r^{2}}b^{\prime}\left(  r\right)  +\frac
{3}{2r^{3}}b\left(  r\right)  \right]  =m_{2}^{2}\left(  r\right)
\end{array}
\right.  , \label{potentials}%
\end{equation}
where we have defined two effective masses dependent on $r$. Now let us
consider the following state%
\begin{equation}
\left\langle t,r,\theta,\phi|\omega,k.l.m\right\rangle =\frac{1}{r}%
f_{i}\left(  x\right)  Y_{lm}\left(  \vartheta,\phi\right)  \exp\left(
-i\omega t\right)  ,
\end{equation}
then the system $\left(  \ref{F p33}\right)  $ becomes%
\begin{equation}
\left\{
\begin{array}
[c]{c}%
\left[  -\frac{d^{2}}{dx^{2}}-\omega^{2}+V_{1}\left(  r\right)  \right]
f_{1}\left(  x\right)  =\tilde{\lambda}_{1,l}^{2}f_{1}\left(  x\right) \\
\\
\left[  -\frac{d^{2}}{dx^{2}}-\omega^{2}+V_{2}\left(  r\right)  \right]
f_{2}\left(  x\right)  =\tilde{\lambda}_{2,l}^{2}f_{2}\left(  x\right)
\end{array}
\right.  . \label{F Lich}%
\end{equation}

\end{document}